\documentstyle[11pt,epsf]{article}

\input psfig
\def\beq{\begin{eqnarray}}
\def\eeq{\end{eqnarray}}
\def\bea{\begin{eqnarray*}}
\def\eea{\end{eqnarray*}}

\def\centeron#1#2{{\setbox0=\hbox{#1}\setbox1=\hbox{#2}\ifdim
\wd1>\wd0\kern.5\wd1\kern-.5\wd0\fi
\copy0\kern-.5\wd0\kern-.5\wd1\copy1\ifdim\wd0>\wd1
\kern.5\wd0\kern-.5\wd1\fi}}
\def\ltap{\;\centeron{\raise.35ex\hbox{$<$}}{\lower.65ex\hbox{$\sim$}}\;}

\def\gtap{\;\centeron{\raise.35ex\hbox{$>$}}{\lower.65ex\hbox{$\sim$}}\;}

\def\singleandthirdspaced{\baselineskip=\normalbaselineskip\multiply
    \baselineskip by 130\divide\baselineskip by 100}
\def\singlespaced{\baselineskip=\normalbaselineskip}

\def\dslash{\not{\hbox{\kern-2pt $\partial$}}}
\def\Dslash{\not{\hbox{\kern-4pt $D$}}}
\def\Oslash{\not{\hbox{\kern-4pt $O$}}}
\def\Qslash{\not{\hbox{\kern-4pt $Q$}}}
\def\pslash{\not{\hbox{\kern-2.3pt $p$}}}
\def\kslash{\not{\hbox{\kern-2.3pt $k$}}}
\def\lslash{\not{\hbox{\kern-2.3pt $l$}}}
\def\qslash{\not{\hbox{\kern-2.3pt $q$}}}
\def\epsilonslash{\not{\hbox{\kern-2.3pt $\epsilon$}}}

\newcommand{\newc}{\newcommand}
\newc{\qbar}{{\overline q}}
\newc{\Kahler}{K\"ahler }
\newc{\deltaGS}{\delta_{\rm GS}}
\begin{document}
\begin{titlepage}
\begin{flushright}
{\large hep-th/0107259 \\ SCIPP-01/27\\
}
\end{flushright}

\vskip 1.2cm

\begin{center}

{\LARGE\bf Dark Matter and Dark Energy:  A Physicist's
Perspective\footnote{Physics summary talk of the conference {\it The
Dark Universe:  Matter, Energy and Gravity}, Space Telescope Institute,
April 2001}}

\vskip 1.4cm

{\large  Michael Dine}
\\
\vskip 0.4cm
{\it Santa Cruz Institute for Particle Physics,
     Santa Cruz CA 95064  } \\

\vskip 4pt

\vskip 1.5cm

\begin{abstract}
For physicists, recent developments in astrophysics and cosmology
present
exciting challenges.  We are conducting ``experiments" in energy
regimes some of which will be probed by accelerators in the near
future, and others which are inevitably the subject of more
speculative theoretical investigations.  Dark matter is an area
where we have hope of making discoveries both with accelerator
experiments
and dedicated searches.
Inflation and dark energy lie in regimes
where presently our only hope for a fundamental understanding lies
in string theory.

\end{abstract}

\end{center}

\vskip 1.0 cm

\end{titlepage}
\setcounter{footnote}{0} \setcounter{page}{2}
\setcounter{section}{0} \setcounter{subsection}{0}
\setcounter{subsubsection}{0}

\singleandthirdspaced


\section{Introduction}

It is a truism that the development of
astronomy, astrophysics, cosmology relies on
our understanding of the relevant laws of physics.
It is thus no surprise that my astronomy colleagues tend to know
more classical mechanics, electricity and magnetism,
atomic and nuclear physics than my
colleagues in particle theory.

As we consider many of the questions which we now face in
cosmology, we must confront the fact that we simply do not know
the relevant laws of nature.  The public often asks us ``what came
before the big bang."  We usually think of this as requiring
understanding of physics at the Planck scale.
But at present we can't even come close.
Ignorance sets in slightly above
nucleosynthesis, and becomes severe by the time we reach the weak
scale.  Some of the questions which trouble us will be settled by
experiment over the next decades; some require new theoretical
developments.  Needless to say, it is possible that much will
remain obscure for a long time.
\begin{itemize}
\item GeV scales:  QCD is by now a well tested theory,
but the phase structure of QCD is not completely
understood, and possible first order phase transitions,
superconducting phases, strange matter, etc. could be relevant
both to astrophysics and cosmology.  These questions may be
settled by improved lattice gauge calculations, and conceivably by
developments at RHIC.
\item  $T=100$ GeV-TeV:  This is the regime of the weak phase
transition.  In order to understand this transition, we need
experimental information on the Higgs particle, or whatever
physics is
responsible for the mass of the $W$ and $Z$.  This physics {\it might}
be
the origin of the matter-antimatter asymmetry\cite{cknreview}.  This is
physics
which will be explored by the Large Hadron Collider at CERN and
by a large electron-positron collider, hopefully to be built by an
international consortium over the next decade.
\item $T = 100$ Gev-TeV:
One of the best-motivated candidates for the dark matter is the
``LSP" (Lightest Supersymmetric Particle) expected if nature is
supersymmetric.  If the supersymmetry hypothesis is correct, we
can expect to encounter this particle and its supersymmetric
cousins at the Tevatron or LHC and the electron-positron collider.
This
physics could well be responsible for the matter-antimatter
asymmetry.
\item $T=1$ Tev: There have been several suggestions over the
past three years that the fundamental scale of physics might
lie at the TeV scale\cite{earlylargedimensions,largedimensions,rs}.
In this case, the Planck scale would be so large --
and gravity so weak -- because there are some very large or highly
warped extra dimensions of space.  If this hypothesis is correct,
there could well be dramatic new phenomena in cosmology
just above the temperature of nucleosynthesis.
\item $T= 10^{10}$ Gev?  The axion is another well-motivated
dark matter candidate.  It is associated with a symmetry known
as a Peccei-Quinn symmetry.  $10^{10}$ GeV might be the scale at
which this symmetry is broken.  It also might be a scale
associated with the dynamics responsible for supersymmetry
breaking.
\item $T=10^{15}$ GeV?  This could well be the scale of the
physics responsible for
inflation.  Over the past few years,
this scale has also emerged as a possible value
for the fundamental scale of $M$ theory\cite{hv}.  The two
might well be connected\cite{banksinflation}.
\item $T > 10^{15}$ GeV?  Perhaps physics at these scales holds the
explanation of the value of the cosmological
constant, and identification of the nature of the dark energy.
Perhaps only here lies the physics which
resolves the singularity of the big
bang, and explains the initial conditions of the universe.
\end{itemize}

For particle physicists it is extremely exciting to think that there are

connections between events in accelerators and our understanding
of the history of the universe.  Perhaps as important, cosmology can
serve as a testing grounds for ideas which are not so
readily studied in more conventional experiments.
We are probably not going to answer all of the questions which I have
listed here soon.
But it {\it is} remarkable that, as we will see, we have hopes of
attacking all of them.

\section{String or M Theory}

I will take string theory as a theoretical umbrella in this talk.
String theory is a natural framework to talk about all of the
issues I have raised above.  Indeed, for many of these questions,
it is the only framework we have.
First, string theory is our only consistent theory of gravity and
quantum
mechanics.  Such a framework is essential if we are to address
many of the questions which we face in cosmology.
Equally important, string theory encompasses virtually every idea we
have for dark matter
and energy:
\begin{itemize}
\item
Low energy supersymmetry, with symmetries like $R$ parity which give
rise to
a stable, weakly interacting particle.
\item
Axions: As I will explain further below, string
theory is the {\it only} theoretical context in which
we can make sense of the axion hypothesis.
\item
Cosmological constant:  String theory is the only theoretical framework
in which
we can, even in principle, calculate the cosmological constant.  It
has realizations all of the various proposed solutions:  it has
candidates for multiple vacua which might produce an anthropic
solution, as mentioned in Vilenkin's talk; it can produce
extremely light particles, which realize the other anthropic
proposal which Vilenkin mentioned\cite{vilenkin}; and
it is ``holographic" (to be explained below), so it might offer entirely

new solutions.
\item
Quintessence:  String theory is the only context in which we can
sensibly discuss the sorts of extremely flat potentials necessary
to realize the ideas of
quintessence\cite{quintessence},\cite{choiquintessence}.
\end{itemize}

What is also striking about string theory is that it will allow us
to make rather definite statements about many of these ideas.   We
will see that within our current understanding of string theory,
the two anthropic solutions which I mentioned above are
implausible.\footnote{This statement requires some qualification; see
\cite{bdm}.}  Physicists tend
to view anthropic explanations of features of physical law with
skepticism
or worse.
Personally, I have for many years thought we might have to
contemplate an anthropic solution of the cosmological constant
problem\cite{weinberglambda,weinberganthropic,vilenkinanthropic}.
I realized at this meeting
that astronomers are more receptive to
these ideas than physicists.  But
what is significant here is that we can potentially use string
theory to rule out some anthropic explanations on {\it scientific}
rather than philosophical grounds.

While I will not stress the point here, by similar reasoning,
the idea of quintessence similarly extremely difficult to realize in
string
theory\cite{choiquintessence}.

\subsection{String Theory:  A Quick Introduction and Survey of
Recent Developments}

What is string theory?  At the most simple level, it is just that:
a theory of quantized strings.  Such a theory is {\it
automatically} a theory which is generally covariant with
non-abelian gauge groups.
Why?  While there has been much progress in understanding these
theories, we have at best only a glimpse as to the answer to this
question.

More generally, string theory is a framework in which we might
hope to address a variety of questions both in particle physics
and cosmology.  While it is often said to be a theory in ten
dimensions, it has solutions with different numbers of dimensions,
including four, and
\begin{itemize}
\item
Standard model gauge interactions
\item
Repetitive generations (e.g. 3) of quarks and leptons
\item
Low energy supersymmetry
\item
Discrete symmetries (R-parity)
\item
Axions
\item
Light scalars with very flat potentials (inflatons? quintessence?)
\item
Exotic possibilities, such as large ``compact" dimensions, with
dramatic possible implications for particle physics, astrophysics
and cosmology.
\end{itemize}

In the last few years, there have been a number of developments,
which are usually grouped together under the heading of duality:
\begin{itemize}
\item
What were once thought to be several independent string theories have
been
recognized to be states of one large theory (sometimes
called $M$ theory)\cite{wittenusc}.
Given that the difficulties of quantizing gravity seem so
immense, the fact that all previously successful attempts
are part of one structure suggests that,
just as there is a unique theory of fundamental vector bosons
interacting with matter, so there may truly be a unique theory of
gravity.
\item
Many interesting dualities have been understood.  For example, many
string (and
field) theories exhibit an exact electric-magnetic duality (see
Jackson, chapter 6!)
\item
Many new theoretical tools have been developed, which have permitted
the study of quantum aspects of black holes and other real phenomena
of quantum gravity.  For example, the Beckenstein-Hawking entropy
has been understood through the counting of microscopic
states\cite{stromingervafa}.
\item
A striking new principle of quantum gravity has been discovered,
known as the
Holographic principle\cite{holography}:  quantum theories of gravity
have far fewer degrees of freedom than conventional quantum
field theories, such as those of the Standard Model.
The number grows like the surface area of the system
rather than the volume.  This is likely to have profound
consequences for the understanding of the question of the
cosmological constant\cite{littlelambda,kaplannelson,thomas}
and other issues in cosmology\cite{banksfischler}.
\end{itemize}

These developments have provided a number of new insights into
longstanding problems.  For example,
we used to think that the basic scales in string theory would be
of order the Planck scale.  But with the new developments,
we have recognized new possibilities.
\begin{itemize}
\item
In the strong coupling limit, string theory
is best described in terms of an eleven dimensional theory, with
gravity propagating in all eleven dimensions, while gauge interactions
are confined to ten dimensional walls\cite{hv}.
There is some evidence that this limit is the best suited for
describing the real world\cite{wittency}.  If this idea is correct,
the fundamental scale of this theory, the eleven dimensional
Planck mass, satisfies:
\beq
M_{11} \sim 10^{15} {\rm GeV}
\eeq
The eleventh dimension is curled up, along with the other (more
conventional(!?)) six, with radii $R_{11}$ and $R$ given roughly
by:
\beq
R_{11}M_{11} \sim 10-30  ~~~~~~ R^6 M_{11}^6 \sim 60
\eeq
The values of $G_N$, and the unification of the gauge couplings,
give support for this picture.
\item
Traditionally, in thinking about compactification, one imagined
that any extra dimensions were extremely small, of order the
Planck mass or unification scale.  In recent years, it has been
appreciated that extra dimensions might be far
larger\cite{largedimensions}, or could be highly curved\cite{rs}.
Either possibility, it has been suggested, might provide an
alternative to supersymmetry as a solution to the hierarchy
problem.  (Prior to these developments, while large extra
dimensions had occasionally been suggested, it had not been
possible to make sense of them.)
These new proposals involve walls or branes in a crucial
way, much as in the eleven dimensional limit.
The fundamental scale of physics lies at $1$
TeV, or so; the smallness of Newton's constant is due to the large size
of the extra dimensions, through the relation:
\beq
G_N = M_{{fund}}^8 V_{{comp}} ~~~M_{{fund}}
\approx {\rm TeV}.
\eeq
\end{itemize}

The eleven dimensional picture predicts low energy supersymmetry,
but also possesses scales of the sort needed
to understand the features of inflation\cite{banksinflation}.  The
large dimension idea predicts:
\begin{itemize}
\item  Dramatic growth of cross sections for production of
Kaluza-Klein modes (e.g. in the process $e^+ + e^- \rightarrow
\gamma + {\rm missing ~ energy}$).
\item  Cosmology:  effects of Kaluza-Klein modes might be
important just above nucleosynthesis.  For example,
some dimensions might be much
smaller at early times.
\end{itemize}

\subsection{What Makes String Theory Hard?}

What makes string theory hard?  Why don't we have all the
answers?  Part of the answer is simply that it is an
ambitious theory.  It's supposed to explain all the facts of
the standard model, {\it with no parameters}.
It is not reasonable to expect all of the answers to fall out so
easily.  But there are also some specific problems:
\begin{itemize}
\item
While I said that there are states with
desirable properties, there are in some sense
too many states.  For example, there are states with
11,10,9,8,...4,3 dimensions; states with or without supersymmetry;
states with 1-100's of generations, and so on.
\item
The classical solutions possess continuous parameters.  From the
perspective of ``low energy"
physicists, these are associated with fields.  These fields are called
moduli.
Examples include a field called the dilaton, whose expectation value
determines the values of the gauge couplings;, and the radius.
We will denote these by  $g^{-2}(x^{\mu})$ (``dilaton") and
$R^2(x^\mu)$,
respectively.
\item
Cosmology:  Moduli are candidates for the inflaton,
but they also lead to a set of cosmological
difficulties.
\item
Cosmological constant (more later)
\end{itemize}

\section{DARK MATTER}

Particle physics has provided at least two plausible
candidates for dark matter.
\begin{itemize}
\item
The lightest supersymmetric particle (LSP):  requiring that the proton
lifetime be long
in supersymmetric theories
almost inevitably means that the LSP is stable.  Supposing that
supersymmetry is broken at a scale of order 1 TeV,
automatically leads to a relic density for this particle of
roughly the right order to be the
dark matter\cite{darkmatter}.
%

In supersymmetric field theories, it is necessary to postulate discrete
symmetries from
nowhere to explain the stability of the proton; in string theory,
such symmetries are ubiquitous\cite{gsw}.  We heard at this meeting
descriptions of ongoing searches for these particles.  While the
hints from the DARMA experiment are controversial,  the $2.6
\sigma$ discrepancy in $(g-2)_{\mu}$ provides some cause for
optimism that direct evidence for supersymmetry will soon be
found\cite{g-2}.  Indeed, a number of physicists have argued for some
time
that if the supersymmetry hypothesis is correct, one is likely to
see a discrepancy in $g-2$\cite{marciano}.  Over the next year, further
data will
be analyzed and the error bars will shrink significantly.
\item
Axions:  The axion is associated with strong CP
problem\cite{axionreview}.  The
axion idea predates the realization that string theory
possesses axions by several
years\cite{wittenaxion},
but it is in string theory that the idea finds a natural
home, and indeed it would inevitably would have been discovered there
had
it not been suggested earlier.  In field theory, the Peccei-Quinn
solution of the strong CP problem requires that one
postulate that nature has a symmetry, which is
broken only by tiny quantum effects in the strong
interactions.  This symmetry  must hold so accurately
that {\it extremely} tiny gravitational effects would spoil it, and
it has sometimes been argued that this is
implausible\cite{mrkamionkowski}.
But in string theory, Peccei-Quinn symmetries of exactly the
desired type automatically arise.
Prior to the understanding of duality, the Peccei-Quinn scale in string
theory was most
naturally identified as $M_p$, so if the
axions constituted the dark matter, they were undetectable.  With the
new
understanding, many other possibilities have emerged\cite{bdaxions}.
\end{itemize}


\section{The Problem of the Dark Energy}

As we have heard at this meeting, the evidence for dark energy is
mounting.  As Professor Livio stressed in his summary, a year ago
many astronomers would have doubted the existence of dark energy.
Now most, if not totally convinced, are starting to believe it.
As Professor Perlmutter remarked, it is particle physicists,
especially theorists, who have been his biggest skeptics.  As we
will see, this is because the result is so surprising.  But given
that it now seems likely that the data -- and its interpretation
as dark energy -- are correct, it is necessarily
a profound clue to the nature of physics at
some very different scale.


>From the perspective of a particle theorist, the question is:
what is the energy density of the vacuum, i.e. of the ground state
of whatever is the underlying theory of nature.  Obviously it is a
tall order to compute this -- we need to know the theory -- but
dimensional analysis suggests we are in trouble.  Particle
physicists like to describe the cosmological constant, $\Lambda$, as a
quantity of dimensions of $[{\rm mass}]^4$.  So
\beq
\Lambda = M^4
\eeq
where $M$ should be some characteristic mass scale in physics.
Is
$M= M_p$ (the Planck mass)?  $M=M_Z$? $M=m_p$ (the proton mass)?
Even in the last case, we would be off by $47$ orders of
magnitude!
Could there be some principle which simply predicts zero?  If so,
what is the origin of the very tiny observed value?

In string theory, there is some good news with regards
to this problem.  At the classical level, all of the string vacua
I have mentioned have vanishing cosmological constant.  While
technically easy to describe\cite{gsw}, this fact is in many ways
mysterious.  It is not a consequence of symmetries of
space time.  So this fact represents
a striking failure of dimensional analysis, of just the sort
we want!

However, even if $\Lambda=0$ at the classical level of some theory, it
is
very hard to understand why quantum effects wouldn't generate a
huge value for it.
The quantum theories of the standard
model describe approximately free fields, i.e they are collections of
harmonic oscillators, one for each momentum, spin (and other quantum
numbers).  The ground state energy
of such a theory is then, to lowest order in $\hbar$,
\beq
\Lambda =\sum (-1)^F \int {d^3 k \over (2 \pi^3)} \sqrt{\vec k^2 +
m^2}.
\label{vacuumenergy}
\eeq

In general, this expression is very divergent.  In our
understanding
of effective quantum field theories, so successful in describing
the standard model, this means that $\Lambda$ is
just a parameter of the theory.  This is why physicists (with a
few exceptions)
traditionally ignored this problem.

On the other hand, at some level, if this is the correct way to
think about the vacuum energy, some physics must cut off this
integral.  What might this be?
If nature were exactly
supersymmetric, then the bosonic and fermionic contributions to
$\Lambda$ would cancel.  If nature is approximately
supersymmetric, the integral diverges quadratically, and assuming that
physics at the Planck scale provides the cutoff,
\beq
\Lambda \sim  M_{susy}^2 M_p^2 \approx (10^{10} {\rm GeV})^4?
\eeq


In string theory, there are no divergences.  Since this is a
theory of gravity, the calculation of the cosmological constant
should be well defined.
So this should be a good test of string theory.
Does it pass?  Is $\Lambda = 0$?  $10^{-47}$

Here we have the not so good news:
We don't know the answer to this question.
All of the string vacua we understand possess moduli at the classical
level.  These are the light
fields with no potential which I referred to above.  The expectation
values
of these fields determine the coupling constant of the string theory,
and the
masses of various states. Often we can
compute a potential for these moduli, but the potential always
tends to zero as the coupling tends to zero.
Indeed, we have no examples where we can find a stable minimum of the
potential in a completely controlled approximation, since, almost
by definition, our approximations break down at such a point.



When one discusses moduli, with potentials which fall to zero at
infinity, it is natural to consider quintessence.
So far, I have spoken of the dark energy as a cosmological
constant.  As we have heard,
many authors have considered the possibility that the dark energy
represents some form of quintessence, which I will loosely
refer to as the energy of some
time-varying field.  This is an interesting idea, if only as
the equation of state for such a field provides a measure of the
quality of future experiments to study the dark energy.   It
should be noted that, whatever the details of the underlying
theory, the mass of the quintessence field today (the second
derivative of its potential) can not be significantly smaller than
the current horizon size.  This is an extremely small number in
particle physics units.  In other words, not only must the actual
value of the present energy density be extremely tiny but so must
other quantities.

In string theory, however, it is hard to make sense of the
quintessence idea,
precisely for the reasons I gave above.  The
difficulty is that, in examples we can analyze, the scale of
the potential is connected to the scale of supersymmetry
breaking, which is much too large.  So in some sense, one needs to
fine tune not only the scale of the potential, but also its
derivatives, with extreme precision\cite{choiquintessence}.  (Since this

talk was
presented, two papers have appeared noting that there are also
serious conceptual issues with quintessence in string
theory\cite{fischleretal,susskindetal}.)  Quintessence also does
not provide a simple explanation of the ``why now" puzzle:  the
question why the cosmological constant, now, is comparable to the
energy density of dark matter.  For example, in a simple
model\cite{albrecht}, this question is resolved by fine-tuning an
additional parameter, at the percent or fraction of a percent
level.  This is not to say that observers should not focus their
efforts on measuring $w$.  At the very least, such measurements
will give us further confidence that there is a large dark energy
component.

Given that quintessence does not fit easily into our current
understanding of string theory, let us return to the more
conventional cosmological constant idea.  If there are stable minima,
they are in regions where we
can't calculate.  Do such stable states exist? Does
the cosmological constant vanish, or is it very small for such states
(as a consequence of some principle)?
Might there be many such states so that we could implement an anthropic
solution?

Here there is (reasonably) good news:
\begin{itemize}
\item
String theory possesses features which allow us to discuss
anthropic solutions of the cosmological constant of the type discussed
by
Vilenkin at this meeting.  It has been argued that there might be a very

dense set of
vacuum states\cite{polchinskietal,wilczeketal}.
There can be very flat potentials.
\item
String theory is {\it not} like field theory.  There is good
evidence that it does not possess nearly as many degrees of
freedom as field theory.  So perhaps the naive quantum estimate
we described above is not correct.  We might then hope that the
classical
cosmological constant vanishes, and that the quantum contributions
are much smaller than naively expected.
\end{itemize}

It is interesting that both of theses ideas suggest that the
cosmological constant is very small, but not zero.

\section{The Anthropic Principle}

To the ``why now" question, a number of answers have been offered.
Through the years, many authors have noted that an energy scale of
order $1 {\rm TeV}$ is a natural scale to consider in physics, and
that $G_N^2 [{\rm TeV}]^8$ is within an order of magnitude of the
observed dark energy (in the past, it was argued that it was
within such a factor of the limit on the cosmological constant).
This is quite impressive, until one remembers that we are indeed
trying to explain a coincidence within a factor of two, and that
the choice of $TeV$ is very rough.  E.g. if it happened that the
correct scale was $3$ TeV, we would be off by nearly $10^4$!

Unfortunately, one can't help but look at the data and conclude
that it is pointing us in the direction of some sort of anthropic
explanation.  Perhaps, if the cosmological constant were much
different than observed, the conditions for life, even in its most
rudimentary conceivable form, might not be satisfied?  I observed
at this meeting that astronomers are less afraid to contemplate
such a prospect than physicists; they are aware of numerous
coincidences in nature which may require such an explanation.  In
the company of many of my physics colleagues, mentioning the
cosmological constant is viewed as barely better than advocating
creationism.  As Weinberg has remarked:
A physicist talking about the anthropic principle runs the
same risk as a cleric talking about pornography:  no matter how
much you say you're against it, some people will think you're a
little too interested.

This topic has been reviewed by Vilenkin at this meeting, and I will
only add a few remarks.  What I have in mind by the anthropic principle
in this context is what Weinberg calls the ``Weak anthropic principle."
The idea is that the universe is vastly larger than what we see within
our horizon.  In different regions of the universe, the cosmological
constant,
and possibly other physical constants, take different values.
Then just as people can only live on planets with water, atmospheres,
etc. (or, just as fish can only live in water),
galaxies/stars/planets/people can only exist in a tiny fraction of
the full universe.  From galaxy
formation, it was originally argued
that this hypothesis could not explain a cosmological constant as
small as observed\cite{weinberglambda,banksunpublished}.  More refined
arguments give results which may be compatible with what we
see\cite{vilenkin}.

As Vilenkin described, there have been a variety of proposals as to
how the laws of nature might admit such variation of the
parameters.  One possibility is that the system has a huge (discrete)
number
of possible (metastable) ground states, and the distribution of
the corresponding energies is nearly
continuous\cite{bds,polchinskietal,wilczeketal}.  This quasicontinuous
distribution
of states has been dubbed a ``discretum"\cite{polchinskietal}.  Note
that the
number of states must be enormous.  If, for example, the typical scale
of the
energies is of order $1$ TeV, then the number of states must be at least

of order
$10^{61}$.  A second
possibility is that the universe is permeated by an extremely
light field, with Compton wavelength large compared to the present
horizon.  As a result this field is currently frozen, but during
the inflationary era, it fluctuated over a range of values,
large enough that in some regions
it cancelled any preexisting cosmological constant.

One may not find this mode of explanation appealing, but in some
sense it may not matter.   Within
string theory, one can argue that neither of these proposed
explanations is very plausible.  Consider, first, the possibility of
a very light field, $\phi$.  The mass of this field has to be smaller
than
$10^{-50}$ GeV.  There are mechanisms in string theory which could
produce a particle this light.  But these mechanisms all imply
that the maximal value of the field is of order $M_p$.
But in order to cancel off a
cosmological constant of order $10^{12} GeV^4$, we need
$\phi \sim 10^{40} M_p$ or so\cite{smatrix}!

In \cite{bds}, it was argued that a peculiar type of axion, known
as the ``irrational axion," might
give rise to a suitable discretum, but subsequent searches have
failed to turn up any examples of the required phenomenon in
string theory.
Four form fluxes in string theory might provide the
necessary ``discretum" to understand the cosmological
constant\cite{polchinskietal,
wilczeketal}.
Whether this works in detail requires resolving many difficult
questions.  For example, we need to understand the stability not
of just one but of $10^{120}$ (or so)states.  It also
raises the specter
that all quantities (the gauge coupling constants, the masses of
the elementary particles....) would all be determined
anthropically (or alternatively would be random numbers)\cite{bdm}.
It is hard
to imagine that all of the standard model parameters are anthropic.
Nor do they look like random numbers.  So while this idea is the
most difficult of the set to rule out, it does not see
particularly promising.

In sum, the remarkable coincidence of the cosmological constant
and the present dark matter density is very suggestive of an
anthropic explanation.  But an anthropic explanation, to be
scientific, requires a sensible underlying theory, presumably in the
context
of a theory which is capable of making other predictions.  So far,
we don't have such a theory.

\section{The Holographic Principle}

't Hooft and Susskind argued, from considerations of black hole
physics, that in a sensible theory of gravity, in
a region of volume V and surface area $A$, the
number of degrees of freedom must be proportional to
$A$\cite{holography}.  The
most familiar piece of evidence for this is the
Beckenstein-Hawking entropy formula:
\beq
S = {G_N A \over 4}.
\eeq
Other features of black hole physics also support this.
The fact that in some sense the information about what is going on
in a large volume is encoded in degrees of freedom residing on the
surface is the origin of the term holographic.

There is some evidence that string theory is holographic:
\begin{itemize}
\item  Naive notions about numbers of degrees of freedom are not
correct in string theory.  For example, string ground states in
smaller dimensions of space time have more degrees of freedom
(suggesting that compactified theories are more ``fundamental"
than uncompactified ones).
\item
String perturbation theory has holographic features:  the $S$
matrix seems to be the crucial observable; C. Thorn argued long
ago that the perturbation theory itself has one the degrees of
freedom of a theory in $d-1$ dimensions.
\item
Two non-perturbative formulations of string theory are known (the
``Matrix Model" and the AdS-CFT correspondence).  Both are
explicitly holographic.
\end{itemize}

What might be the implications of this principle for the Cosmological
Constant?
These are not clear, but they seem likely to be dramatic, since in
the cosmological constant expression, eqn. \ref{vacuumenergy}, one might

no
longer have $V\int d^3 k$, but instead a sum over far fewer
degrees of freedom.
Are there few enough?  The problem is not
sufficiently well understood to say at the present
time\cite{kaplannelson}.

This sort of reasoning has lead to even more radical conjectures.
In string theory, there seem to be states with varying
numbers of dimensions, and varying amounts of supersymmetry.
Many states with unbroken supersymmetry can be argued to be exact
solutions of the theory.
Susskind has suggested that perhaps De Sitter space is not allowed in
string
theory.  He offers no solid argument, but points to
hints based on holography.
Banks proposes that we think
very differently about the question of supersymmetry
breaking\cite{littlelambda}.
He
argues that the number of states
in De Sitter space is finite.   Given that recent observations suggest
that our universe is De Sitter.  What determines the number of
states?  Banks proposes that this number is a {\it parameter}.
The cosmological
constant and the amount of supersymmetry breaking are determined
by this parameter!
This proposal explains why states with too much supersymmetry
might
not be viable (and it is the only proposal which does so).
It requires that in holographic theories a different relation
between $\Lambda$ and the scale of
supersymmetry breaking holds.  There is some
reason to believe this might be the case.  It requires, however, a
very surprising relation to hold between the cosmological constant
and the scale of supersymmetry breaking.

Both ideas are highly speculative, and they are not (yet) supported by
a substantial amount of evidence.  But they are suggestive.
Indeed, if nothing else, they indicate the sorts of radical
rethinking of many of our basic ideas in physics which may be
required to understand the dark energy.

\section{Conclusions}

Particle physicists are eager to know the answers to the
questions:
\begin{itemize}
\item
What is the dark matter?
\item
Is there really dark energy?  What is its equation of state?
\end{itemize}

My experimental colleagues are very interested in dark matter
searches, SNAP and ground based proposals to study Type Ia
Supernovae.   Theorists are hopeful that they have predicted the
correct form of the dark matter; they are frantically trying to
explain the dark energy.  Both are sure to lead to important
insights into fundamental law.

\subsection{Acknowledgments}

I would like to thank many of my colleagues for discussions, and
for educating me about the issues raised here,
especially Tom Banks, Nathan Seiberg and Leonard Susskind.  I would like

to thank
the organizers and participants in this conference for a very
stimulating experience, and particularly to thank Mario Livio.
This work was supported in part by the U.S. Department of energy.




\begin{thebibliography}{99}
\bibitem{darkmatter} The proceedings of this conference feature
up to date reviews of both theoretical and experimental issues
of dark matter and its detection.
\end{thebibliography}
\end{document}